\documentclass[12pt]{article}
\usepackage{graphicx}
\usepackage{amssymb}
\usepackage{amsmath}
\numberwithin{equation}{section}

\def\BR{{\mathbb R}}

\newtheorem{Pa}{Paper}[section]
\newtheorem{Tm}[Pa]{{\bf Theorem}}
\newtheorem{La}[Pa]{{\bf Lemma}}
\newtheorem{Cy}[Pa]{{\bf Corollary}}
\newtheorem{Rk}[Pa]{{\bf Remark}}

\newtheorem{Dn}[Pa]{{\bf Definition}}
\newtheorem{Pn}[Pa]{{\bf Proposition}}

\title{3D Schr\"odinger equation: scattering operator, scattering amplitude and ergodic property}
\author{Lev Sakhnovich}

\date{}

\begin{document}
\maketitle

\emph{99 Cove ave., Milford, CT, 06461, USA}

\vspace{0.3em}
 
 E-mail: lsakhnovich@gmail.com
 
 \vspace{0.3em}

 

\begin{abstract}
Stationary scattering problem (when the distance $r$ tends to infinity)
and dynamical scattering problem  (when the time $t$ tends to infinity)
are considered for the 3D Schr\"odinger equation.
A simple interconnection  between the scattering amplitude (stationary case) and
scattering operator (dynamical case) is given in the paper.
This result  is a quantum mechanical
analog of  the ergodic formulas in the classical  mechanics.\end{abstract}

\section{Introduction}
In the present paper, we consider the Schr\"odinger operator
$\mathcal{L}u(r)=-\Delta{u}+V(r)u,$ where $\Delta$  denotes the 3-dimensional Laplacian,
$V(r)=\overline{V(r)}$ and $r{\in}\BR^{3}.$  The Schr\"odinger operator $\mathcal{L}$
is one of the most important operators in mathematics and physics. Spectral and scattering problems
for this operator were investigated by many authors (see, e.g. important works \cite{IK, Kato, LL, MM, Pov},\cite{MM}).

In this paper we consider two types of scattering problems: stationary scattering problem (when the distance $r$ tends to infinity)
and dynamical scattering problem  (when the time $t$ tends to infinity).

The connection between the scattering amplitude (stationary scattering problem) and
dynamical scattering operator is described by formula \eqref{3.11} and presents the main result
of the paper. Formula \eqref{3.11} is a quantum mechanical
analog of  the ergodic formulas in the classical  mechanics. Works by Povzner \cite{Pov}, Ikebe \cite{IK}
and Kato \cite{Kato} were essential for  our research. 

For the radial case, where $V(r)=V(|r|),$
the quantum  mechanical
analogs of  the ergodic formulas in classical  mechanics were obtained in our papers \cite{Sakh7, Sakh8}.
Some ergodic results were obtained for the Dirac equation as well (see \cite{Sakh9, Sakh10}).
 \section{Preliminary results}
 In this section, we present the results that we
will be needed later.\\
1. Let us consider the Schr$\ddot{o}$dinger operator
\begin{equation}\mathcal{L}u(r)=-\Delta{u}+V(r)u,\quad \mathcal{L}_{0}u(r)=-\Delta{u},\label{2.1}\end{equation}
where $r=(r_1,r_2,r_3){\in}{\BR}^3$.
Further we assume that
\begin{equation}V(r)=\overline{V(r)}.\label{2.2}\end{equation}
Let us write the Lippmann- Schwinger equation:
\begin{equation}\phi(r,k)=e^{ik{\cdot}r}-\frac{1}{4\pi}\int_{\BR^{3}}\frac{e^{i|k||r-s|}}{|r-s|}V(s)\phi(s,k)ds,
\label{2.3}\end{equation}where $|k|=\sqrt{\lambda},\, \lambda>0.$
We note that the solution $\phi(r,k)$ of the Lippmann- Schwinger equation \eqref{2.3} satisfies the relation
\begin{equation}(\mathcal{L}-\lambda)\phi=\lambda\phi.\label{2.4}\end{equation}
The modified Lippmann-Schwinger equation has the form
\begin{equation}(I+K(\lambda))\psi(r,k)=e^{ik{\cdot}r}|V(r)|^{1/2},\label{2.5}\end{equation}
where the operator $K(\lambda)$ is defined by the relation
\begin{equation}K(\lambda)f(r)
=\frac{1}{4\pi}\int_{\BR^{3}}|V(r)|^{1/2}\frac{e^{i|k||r-s|}}{|r-s|}W(s)|V(s)|^{1/2}f(s)ds.\label{2.6}\end{equation}
2. We need the following definition:
\begin{Dn}\label{Definition 2.1}The potential $V(r)$ belongs to Rollnik class if the operator
\begin{equation}A(\lambda)=\int_{\BR^{3}}|V(r)|^{1/2}\frac{e^{i|k||r-s|}}{|r-s|}|V(s)|^{1/2}f(s)ds.
\label{2.7}\end{equation} belongs to the Hilbert-Schmidt class.\end{Dn}
\begin{Rk} \label{Remark 2.2} Further we assume that the potential $V(r)$ belongs to the Rollnik class.
Hence the operator $K(\lambda)$  belongs to the Hilbert-Schmidt class.\end{Rk}
\begin{Dn}\label{Definition 2.3}The point $\lambda>0$ is an exceptional value if the equation $[I+K(\lambda)]\psi=0$ has nontrivial solution in the space $L^2({\BR}^3)$.\end{Dn}
 We denote by $\mathcal{E}_{+}$ the set of exceptional points and we denote by $E_{+}$ the set
 of such points $\lambda>0$ that  $\lambda{\notin}\mathcal{E}_{+}.$\\
 \begin{La}\label{Lemma 2.4} If $\lambda{\in}E_{+}$, then equation \eqref{2.3} has one and only one
solution $\psi(r,k)$ in $L^2({\BR}^3)$. \end{La}
\begin{Cy}\label{Corollary 2.5} If $\lambda{\in}E_{+},$
 then equation \eqref{2.1} has one and only one
solution $\phi(r,k)$ which satisfies the condition $|V(r)|^{1/2}\phi(r,k){\in}L^2({\BR}^3)$. \end{Cy}
We formulate the following result (see \cite{RS}, p.115).
 \begin{Cy}\label{Corollary 2.6}
1) The operator $\mathcal{L}$ has only discrete spectrum in the domain $\lambda<0$.\\
2) The discrete spectrum  of the  operator $\mathcal{L}$ has no limit points in the  domain $\lambda<0$.\\
3)The set $\mathcal{E}_{+}$ is  bounded,  closed and has  Lebesgue  measure equal to zero. \end{Cy}
Let us consider the scattering operator $S(\lambda)$ in the energetic representation. It is
proved (see \cite {RS}, p.110) that $S(\lambda)$ is unitary. Let us introduce the operator function
\begin{equation}T(\lambda)=(2i\pi)^{-1}[I-S(\lambda)]. \label{2.8}\end{equation}
We need the following assertion (\cite{Kato})
\begin{Pn}\label{Proposition 2.7} For each
$\lambda{\in}E_{+}$  the operator $T(\lambda)$    can be represented in the form
\begin{equation}T(\lambda)=\mathcal{F}(\lambda)W(r)[I+K(\lambda)]^{-1}
\mathcal{F}^{\star}(\lambda)\label{2.9}\end{equation}\end{Pn}
 The operator $\mathcal{F}(\lambda)$ is defined by the formula
\begin{equation}\mathcal{F}(\lambda)f(r)=\mu\int_{{\BR}^3}
e^{-i|k|\omega{\cdot}r}|V(r)|^{1/2}f(r)dr,\quad |k|^2=\lambda,
\label{2.10}\end{equation}
where $\mu=(1/4)(\lambda)^{1/4}\pi^{-3/2}$.
The adjoint to $\mathcal{F}(\lambda)$ operator has the form
\begin{equation}\mathcal{F}^{\star}(\lambda)h(\omega)=\mu\int_{S^2}e^{i|k|\omega{\cdot}r}|V(r)|^{1/2}h(\omega)
d\Omega(\omega)
\label{2.11}\end{equation}Here by $S^2$ we denote the surface $|\omega^{\prime}|=1$ in the space ${\BR}^3$, $d\Omega$ is the  standard measure on the
surface $S^2$. Now we formulate the Povzner-Ikebe result (see \cite{Pov} and \cite{IK}):
\begin{Tm}\label{Theorem 2.8}Let the function $V(r)$ belong to space $L^{2}({\BR}^3).$ 
 If
\begin{equation}|V(r)|=O(|r|^{-3-\delta}),\quad \delta>0,\quad |r|{\to}\infty,\label{2.12}\end{equation}
then the solution $\phi(r,k)$ of Lippmann-Schwinger equation \eqref{2.3} has the form
\begin{equation}\phi(r,k)=e^{ik{\cdot}r}+
\frac{e^{i\sqrt{\lambda}|r|}}{|r|}f(\omega,\omega^{\prime},\lambda)
+o(1/|r|),\quad |r|{\to}\infty , \label{2.13}\end{equation}
where $\omega=r/|r|,\,\omega^{\prime}=k/|k|$ and
\begin{equation}f(\omega,\omega^{\prime},\lambda)=-\frac{1}{4\pi}\int_{{\BR}^3}e^{-i\sqrt{\lambda}(s{\cdot}\omega)}
V(s)\phi(s,k)ds.  \label{2.14}\end{equation}
\end{Tm}
The function $f(\omega,\omega^{\prime},\lambda)$ is  the scattering amplitude.\\
\begin{Rk}\label{Remark 2.9}If the conditions of Theorem 2.8 are fulfilled, then
$V(r){\in}L^{3/2}({\BR}^3).$  It follows from the theorem Kato \cite{Kat} that $V(r)$ belongs to the Rollnik class.\end{Rk}
\begin{Rk}\label{Remark 2.10}The following assertion is valid \cite{IK}:\\
If the conditions of Theorem 2.8 are fulfilled, then $E_{+}=(0,\infty)$.\end{Rk}

\section{Scattering operator and scattering amplitude, ergodic property}
1. The scattering operator $S(\lambda)$ is the solution of the dynamical scattering problem, when time $t{\to}\infty$
\cite{IK},\cite{Kato}. The scattering amplitude  $f(\omega,\omega^{\prime},\lambda)$ is the solution of the stationary scattering problem, when distance $r{\to}\infty$. In this section we show that dynamical and stationary scattering problems are closely connected.\\
We begin with the statement.
\begin{Tm}\label{Theorem 3.1}Let $V(r)$ belong to the Rollnik class. Then for each
$\lambda{\in}E_{+}$  the operator $T(\lambda)$  belong to the Hilbetrt- Schmidt class.\end{Tm}
\emph{Proof.} We use the relation
\begin{equation}[\mathcal{F}^{\star}\mathcal{F}f]=\nonumber\end{equation}
\begin{equation}\frac{1}{4\pi^{2}}\int_{{\BR}^3}|V(r)|^{1/2}
\frac{\sin(|k(r-s)|)}{|r-s|}|V(s)|^{1/2}
f(s)ds.\label{3.1}\end{equation}
The  function $V(r)$ belongs to Rollnik class.  Hence  it follows from \eqref {3.1} that the operator
$\mathcal{F}^{\star}\mathcal{F}$ belongs to the Hilbert-Schmidt class. Now using \eqref{2.9} we obtain the assertion of the theorem.\\
Changing the order of integrals in \eqref{2.9} and using \eqref{2.5} we obtain
\begin{equation}T(\lambda)h(\omega^{\prime})={\mu^2}\int_{S^2}T(\omega,\omega^{\prime},\lambda)h(\omega^{\prime})
d\Omega(\omega^{\prime}),
  \label{3.2}\end{equation} where
\begin{equation} T(\omega,\omega^{\prime},\lambda)=\int_{{\BR}^3}e^{-i\sqrt{\lambda}(s{\cdot}\omega)}
V(s)\phi(s,k^{\prime})ds,\label{3.3}\end{equation} and
$ \omega=k/|k|,\, \omega^{\prime}=k^{\prime}/|k^{\prime}|,\,|k|=|k^{\prime}|=\sqrt{\lambda}.$
We note that $\omega$ and $\omega^{\prime}$ are defined in the cases \eqref{2.13} and \eqref{3.3}
differently.
Comparing \eqref{2.14} and \eqref{3.3} we have
\begin{equation}
f(\omega,\omega^{\prime},\lambda)=-\frac{1}{4\pi}T(\omega,\omega^{\prime},\lambda),\quad \lambda>0.
\label{3.4}\end{equation}
2. We introduce the Hilbert space $\mathcal{H}$ of the functions $h(\omega)$. The norm in the space $\mathcal{H}$ is defined by the relation
\begin{equation}\|h\|^{2}=\int_{S^2}|h(\omega)|^{2}d\Omega(\omega).\label{3.5}\end{equation}
We note that the operator $S(\lambda)$ is unitary in the space $\mathcal{H}$. Hence there exists a complete orthonormal system of eigenfunctions $G_{j}(\omega,\lambda)$ of the operator $S(\lambda)$. We denote the corresponding eigenvalues by $\nu_{j}(\lambda)$, where $|\nu_{j}(\lambda)|=1.$
The  function $f(\omega,\omega^{\prime},\lambda)$, where $\omega$ and $\lambda$ are fixed,
 belongs to the space  $\mathcal{H}$. Hence  the  function $f(\omega,\omega^{\prime},\lambda)$ can be represented in the form of series:
 \begin{equation}f(\omega,\omega^{\prime},\lambda)=\sum_{j}a_{j}(\omega)\overline{G_{j}(\omega^{\prime},\lambda)},
 \quad \lambda>0,
 \label{3.6}\end{equation}
where
\begin{equation}a_{j}(\omega,\lambda)=\int_{S^2}f(\omega,\omega^{\prime},\lambda)G_{j}(\omega^{\prime},\lambda)
d\Omega(\omega^{\prime}) \label{3.7}\end{equation}
It follows from \eqref{3.4} and \eqref{3.7}  the relation
\begin{equation}a_{j}(\omega,\lambda)=-\frac{1}{4\pi}\int_{S^2}T(\omega,\omega^{\prime},\lambda)G_{j}(\omega^{\prime},\lambda)
d\Omega(\omega^{\prime}) \label{3.8}\end{equation}
Taking into account \eqref{3.2} and \eqref{3.8} we have
\begin{equation}a_{j}(\omega,\lambda)=-\frac{1}{4\pi\mu^{2}}T(\lambda)G_{j}(\omega,\lambda)
 \label{3.9}\end{equation}
 We recall that $T(\lambda)=[I-S(\lambda)]/(2i\pi)$ and $\mu=(1/4)(\lambda)^{1/4}\pi^{-3/2}$. Hence the relation \eqref{3.9} can be written in the following form:
\begin{equation}a_{j}(\omega,\lambda)=\frac{2\pi}{i\sqrt{\lambda}}[S(\lambda)-I]G_{j}(\omega,\lambda),\quad \lambda>0.
 \label{3.10}\end{equation} Relations \eqref{3.6} and \eqref{3.10} imply the following assertion:
\begin{Tm}\label{Theorem 3.2}Let the conditions of Theorem 2.8 be fulfilled. Then we have
\begin{equation}f(\omega,\omega^{\prime},\lambda)=\frac{2\pi}{i\sqrt{\lambda}}\sum_{j}(\nu_{j}(\lambda)-1)G_{j}(\omega,\lambda)
\overline{G_{j}(\omega^{\prime},\lambda)}, \quad \lambda>0.
\label{3.11}\end{equation}\end{Tm}
Let us define the total cross section
\begin{equation}\sigma(\lambda)=\int_{S^2}\int_{S^2}|f(\omega,\omega^{\prime},\lambda)|^{2}d\Omega(\omega)d\Omega(\omega^{\prime}).
\label{3.12}\end{equation}
\begin{Cy}\label{Corollary 3.3}Let conditions of Theorem 2.8 be fulfilled. Then the total cross section has the form
\begin{equation}\sigma(\lambda)=\frac{4\pi^{2}}{\lambda}\sum_{j}|\nu_{j}(\lambda)-1|^{2}.\label{3.13}\end{equation}\end{Cy}
\begin{Rk}\label{Remark 3.4} It follows from the conditions of Theorem 2.8, that the operator $F^{*}F$
(see \eqref{3.1})
 belongs to the Hilbert-Schmidt class (see \cite{Sim}, Ch.1). Then operator $T(\lambda)$  belongs to the Hilbert-Schmidt class too. Hence  the series \eqref{3.11} and \eqref{3.13} converge.\end{Rk}

\begin{Rk}\label{Remark 3.5}Formulas \eqref{3.11} and \eqref{3.13} give the connections between the stationary scattering results  $(f(\omega,\omega^{\prime},\lambda))$ and the dynamical scattering results  $(G_{j}(\omega,\lambda),\,\mu_{j}(\lambda))$. So, formulas \eqref{3.11} and \eqref{3.13} are quantum mechanical analogues of the ergodic formulas in classical mechanics. For radial case, when $V(r)=V(|r|),$ the
quantum mechanical analogues of the ergodic formulas  were obtained in our papers  \cite{Sakh7},\cite{Sakh8}.\end{Rk}
3. Let ua consider separately the classical case, when
\begin{equation}V(r)=V(|r|),\quad \omega^{\prime}(\theta^{\prime},\phi^{\prime})=\omega^{\prime}(\pi/2,0).
\label{3.14}\end{equation}
In other words, we consider the case when the potential $V(r)$ is spherically symmetric
and the incoming wave has z-direction. If potential  $V(r)$ is spherically symmetric then the corresponding complete orthonormal system of eigenfunctions $\mathcal{Y}_{\ell,m}(\theta,\phi)$ of $S(\lambda)$ have the form  (see \cite{LL}, section 28):
\begin{equation}\mathcal{Y}_{\ell,m}(\theta,\phi)=\mathcal{P}_{\ell,m}(\cos{\theta})\mathcal{F}_{m}(\phi),
\label{3.15}\end{equation}
where
\begin{equation}\mathcal{P}_{\ell,0}(\cos{\theta})=\sqrt{\ell+1/2}P_{\ell}(\cos{\theta}),\quad \ell=0,1,2,...,\label{3.16}\end{equation}
\begin{equation}\mathcal{F}_{m}(\phi)=\frac{1}{\sqrt{2\pi}}e^{im\phi},\quad -\ell{\leq}m{\leq}\ell.
\label{3.17}\end{equation}
Here the functions $P_{\ell}(\cos{\theta})$ are Legendre polynomials.
We do not write formulas
for $\mathcal{P}_{\ell,m}(\cos{\theta}), \, (m{\ne}0)$ since we do not  use  them further.
In case \eqref{3.14} we have
\begin{equation}f(\omega,\omega^{\prime},\lambda)=f(\theta,\phi,\lambda).\label{3.18}\end{equation}
The azimutal rotational symmetry of plane wave and spherical potential ensures that
\begin{equation}f(\theta,\phi,\lambda)=f(\theta,\lambda).\label{3.19}\end{equation}
Taking into account the relations
\begin{equation}\mathcal{P}_{\ell,0}(1)=\sqrt{\ell+1/2},\quad \mathcal{F}_{0}(\phi)=\frac{1}{\sqrt{2\pi}}
\label{3.20}\end{equation} we write equality \eqref{3.11} for case \eqref{3.14}:
\begin{equation}  f(\theta,\lambda)=\frac{1}{2i\sqrt{\lambda}}\sum_{\ell=0}^{\infty}(2\ell+1)[\nu_{\ell}(\lambda)-1]P_{\ell}(\cos{\theta}).
\label{3.21}\end{equation}
Let us write the classical formula (see \cite{LL}, section 122 and \cite{MM}, Ch.II) for case \eqref{3.14}
\begin{equation}  f(\theta,\lambda)=\frac{1}{2i\sqrt{\lambda}}\sum_{\ell=0}^{\infty}(2\ell+1)[S_{\ell}(\lambda)-1]P_{\ell}(\cos{\theta}),
\label{3.22}\end{equation} where $S_{\ell}(\lambda)$ is connected with phase shift $\delta_{\ell}(\lambda)$ by the relation
\begin{equation}S_{\ell}(\lambda)=\exp(\delta_{\ell}(\lambda)).\label{3.23}\end{equation}
Comparing relations \eqref{3.22} and \eqref{3.23} we obtain the assertion:
\begin{Cy}\label{Corollary 3.6} Let conditions of Theorem 2.8 and conditions \eqref{3.14} be fulfilled.
Then\\
1. The equality
\begin{equation}\nu_{\ell}(\lambda)=S_{\ell}(\lambda),\quad \lambda>0\label{3.24}\end{equation}
is valid.\\
2. The classical equation \eqref{3.22}  is the partial case of formula \eqref{3.11}. \end{Cy}
We  proved (\cite{Sakh7}, \cite{Sakh8}) relation \eqref{3.24} before by using ordinary differential equations theory.  The total scattering  cross-section $\sigma(\lambda)$ is defined now by the relation
\begin{equation}\sigma(\lambda)=2\pi\int_{0}^{\pi}|f(\theta,\lambda)|^{2}\sin{\theta}d\theta,\quad \lambda>0.
\label{3.25}\end{equation}
 We note that
\begin{equation}\int_{0}^{\pi}|P_{\ell}(\cos(\theta)|^{2}\sin{\theta}d\theta=\ell/2+1
\label{3.26}\end{equation}
It follows from \eqref{3.22}, \eqref{3.23} and \eqref{3.25}, \eqref{3.26} the well-known formula
\begin{equation}\sigma(\lambda)=\frac{4\pi}{\lambda}\sum_{\ell=0}^{\infty}(2\ell+1)\sin^{2}{\delta_{\ell}}(\lambda),
\quad \lambda>0.
\label{3.27}\end{equation}
\begin{Rk}\label{Remark 3.7} The definitions \eqref{3.12} and \eqref{3.25} of
scattering cross-section $\sigma(\lambda)$ are different. The corresponding results  \eqref{3.13}
and \eqref{3.26} are different too.\end{Rk}

 \end{document}